\documentclass[reprint,amsmath,amssymb,aps,showpacs,floatfix,pre]{revtex4-1}

\usepackage{graphicx}
\usepackage{dcolumn}
\usepackage{bm}
\usepackage{amssymb}
\usepackage{amsmath}

\begin{document}

\preprint{}

\title{Dynamics of two-dimensional dipole systems}

\author{Kenneth I. Golden}
\affiliation{Department of Mathematics and Statistics, Department of Physics, University of Vermont, Burlington, VT 05401 USA}
\author{Gabor J. Kalman}
\affiliation{Department of Physics, Boston College, Chestnut Hill, MA 02467 USA}
\author{Peter Hartmann}
\author{Zolt\'an Donk\'o}
\affiliation{Research Institute for Solid State and Optics, Hungarian Academy of Sciences, POB 49, H-1525 Budapest, Hungary}
\affiliation{Department of Physics, Boston College, Chestnut Hill, MA 02467 USA}

\date{\today}

\begin{abstract}

Using a combined analytical/molecular dynamics (MD) approach, we study the current fluctuation spectra and longitudinal and transverse collective mode dispersions of the classical two-dimensional (point) dipole system (2DDS) characterized by the $\phi _{D}(r)=\mu^2/r^3$ repulsive interaction potential; $\mu$ is the electric dipole strength.  The interest in such two-dimensional dipole systems (2DDS) is twofold. First, the quasi-long range $1/r^{3}$ interaction makes the system a unique classical many body system, with a remarkable collective mode behavior. Second, the system may be a good model for a closely spaced semiconductor electron-hole bilayer, a system that is in the forefront of current experimental interest. The longitudinal collective excitations, which are of primary interest for the liquid phase, are acoustic at long wavelengths. At higher wave numbers and for sufficiently high coupling strength, we observe the formation of a deep minimum in the dispersion curve preceded by a sharp maximum; this is identical to what has been observed in the dispersion of the zero temperature bosonic dipole system, which in turn emulates so called roton-maxon excitation spectrum of the superfluid $^4$He. The analysis we present gives an insight into the emergence of this apparently universal structure, governed by strong correlations. We study both the liquid and the crystalline solid state. We also observe the excitation of combination frequencies, resembling the roton-roton, roton-maxon, etc. structures in $^4$He.
\end{abstract}


\maketitle

\section{Introduction}

This paper addresses the dynamical behavior of two-dimensional classical liquid and solid systems consisting of point electric dipoles in a plane, with the dipole moments oriented perpendicular to the plane. This system is equivalent to an ensemble of point particles with a repulsive $1/r^3$ interaction potential. The interest in such two-dimensional dipole systems (2DDS) is twofold. First, the quasi-long range $1/r^{3}$ interaction makes the collective behavior of the system somewhat similar to that of the one-component plasma with its $1/r$ Coulomb interaction, while, at the same time, the $1/r^{3}$ interaction, with its rapid drop-off at large $r$ and its hard singularity at $r = 0$, resembles the typical interaction potentials in classical liquids. This duality makes the 2DDS a unique classical many body system worthy of detailed exploration. The second aspect that makes the study of the 2DDS timely and important is that it can provide, as we will discuss below, a good model for a semiconductor electron-hole bilayer, a system that is in the forefront of current experimental interest \cite{EHexp}, as well as for colloidal suspensions of super-paramagnetic particles \cite{Keim}.

The existence of bound electron-hole excitons in semiconductors was predicted quite some time ago by Keldysh and co-workers \cite{1,1b} and by Halperin and Rice \cite{2}. Electron-hole bilayers (EHBs) are structures especially well suited to the formation of a stable dipole-like excitonic phase \cite{3,3b,3c}.  In such systems, the charges in the two layers with opposite polarities are physically separated from each other, reducing their recombination rate and for sufficiently small layer separations forming  a bound dipole-like excitonic structure.  The formation of the excitonic/dipole phase in the EHB has been confirmed by recent diffusion and path integral Monte Carlo (MC) simulations \cite{4,5,5b}, by classical MC simulations \cite{6}, and by classical molecular dynamics (MD) simulations \cite{7}.  Fixed-node diffusion MC simulations of the zero-temperature symmetric ($m_e = m_h$, $n_e = n_h$) EHB \cite{4} indicate three phases: the excitonic liquid, the spin-unpolarized Coulomb liquid, and the triangular Wigner crystal. Classical MC simulations of the bipolar bilayer \cite{6} indicate the existence of four phases in the strong coupling regime: the excitonic dipole liquid and solid and the Coulomb liquid and solid phases.  The necessary existence of these four phases was also pointed out in \cite{8}.  Since the excitons are bosons, at low temperatures they may also form a Bose-Einstein condensate \cite{4,9,9b,10,11}, become superfluid \cite{9c}, or possibly a supersolid \cite{8}.

Based on the phase diagrams \cite{4,5}, the closely spaced EHB in its dipole-like excitonic phase can be, in a good approximation, modeled as a 2D monolayer of interacting point electric dipoles.  The model 2D dipole system (2DDS) can be described as a collection of $N$ spinless point dipoles, each of mass $m=m_e +m_h$, occupying the large but bounded area $A$; $n=N/A$ is the average density.  The dipoles are free to move in the $x-y$ plane with dipolar moment oriented in the $z$-direction; the repulsive interaction potential is accordingly given by $\phi _{D} =\mu^2 /r^3$, where $\mu$ is the electric dipole strength.  

The coupling strength of the symmetric ($n_e=n_h=n$, $m_e=m_h=m/2$) EHB at arbitrary degeneracy is characterized by the parameter $\tilde{\Gamma}=e^2 /(a\langle E_{\rm kin} \rangle)$, where $e$ is the electrical charge and $a=\sqrt{1/(\pi n)} $ is the 2D Wigner-Seitz radius. In the high-temperature classical domain, this becomes the customary coupling parameter $\Gamma =\beta e^2/a$ ($\beta =1/k_{\rm B} T)$, while at zero-temperature, it becomes $r_s =a/a_{\rm B}$, ($a_{\rm B} =\hbar^2 /m_{e,h} e^{2}$ is the Bohr radius).  By the same token, the high-temperature classical coupling parameter for the 2DDS is characterized by $\Gamma_{D} =\beta \mu^2 /a^3$, whereas at zero-temperature, the appropriate measure of the 2DDS coupling strength is given by $r_{D} =r_0 /a$, where $r_0 =m \mu^{2} /\hbar^{2}$ is the dipole equivalent of the Bohr radius; note the correspondence $\Gamma_{D} \Leftrightarrow r_{D}$. 

In the case of the symmetric EHB in the zero-temperature quantum domain, the Coulombic and dipole coupling parameters are related to each other by $r_D = 2r_s (d/a)^2$, where $d$ is the spacing between layers. High coupling ($r_D \gg 1$) for point dipoles corresponds to the low-density regime in the closely spaced $(d/a < 1)$ EHB, as dictated by the ordering $(2r_s)^{-1} \ll (d/a)^2 < 1$, or equivalently, $a > d \gg a_{\rm B}/2$. In the high-temperature classical domain, the Coulombic and dipole coupling parameters are related to each other by $\Gamma_{D} = \Gamma (d^{2}/a^{2})$; high coupling ($\Gamma_D \gg 1$) for the classical 2DDS liquid emulating the closely spaced EHB is now dictated by the ordering $ \Gamma^{-1} \ll d^{2} /a^{2} <1$, a condition that is easily met for fixed layer density $n$ at sufficiently low temperatures in the classical regime and/or $d$ sufficiently small. In the present work we focus on the classical 2DDS in the strong coupling regime that includes both the dipole liquid and solid phases. Our preliminary MD study indicates that the classical 2DDS liquid freezes at $\Gamma_D \sim 70$.

In the low temperature regime the 2DDS becomes a 2D bosonic dipole liquid, with superfluid properties. The collective mode spectrum of this model has been considered by a number of investigators \cite{12,12b,13,14,19}. Quantum Monte Carlo (QMC) simulations carried out by Astrakharchik and co-workers \cite{12,12b} paved the way by generating essential information about the 2DDS ground-state energy, static structure function $S(k)$, and the 2DDS liquid-solid phase transition, which they estimate to occur at $r_D \sim 30$. Invoking the zero temperature Feynman Ansatz    
\begin{equation} 
\label{eq:1} 
\omega(k) = \frac{\hbar k^{2} }{2mS(k)},  
\end{equation} 
they established an upper-bound estimate of the collective mode dispersion with the input of their MC-generated $S(k)$ data.  Subsequently, Mazzanti and co-workers \cite{15} generated collective mode spectra based on the more sophisticated  correlated basis functions (CBF) formalism \cite{16,17}, using $S(k)$ as an input. 

\begin{figure}[htb]
\includegraphics[width=0.9\columnwidth]{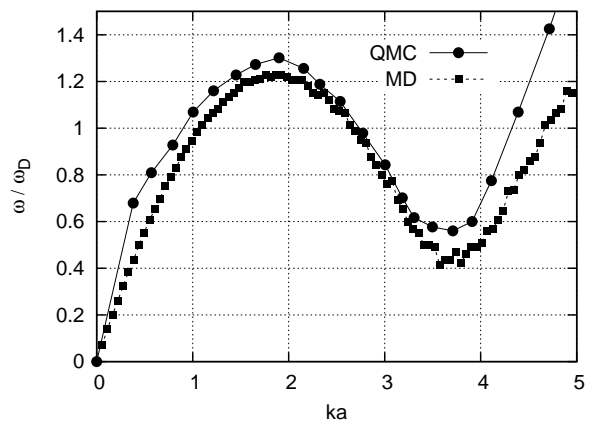}
\caption{\label{fig:1} 
Longitudinal collective mode dispersion curves generated from classical molecular dynamics (MD) simulations (squares) at $\Gamma _{D} =28$ and from the Ref. \cite{12,12b} zero-temperature quantum Monte Carlo (QMC) simulations (dots) at $r_{D} =28.4$.  $\omega _{D} =\sqrt{2\pi n\mu ^{2} /ma^{3}}$ is a characteristic dipole oscillation frequency.}
\end{figure}

In contrast to the above studies that require additional assumptions to make it possible to infer information about the excitation spectrum from static structure function data, an entirely different approach, namely classical MD simulations can provide direct insight into the structure of the collective modes. We argue that such information, even though based on classical dynamics, is pertinent both to the classical and to the quantum domains: it is expected that the nature of the collective excitations would not be all that different in the two domains. This is borne out by our preliminary MD studies \cite{18,19} of the longitudinal collective mode dispersion in the strongly coupled classical 2DDS liquid: we have observed that the classical dispersion reproduces the qualitative features of the collective mode dispersion calculated from the Feynman formula \eqref{eq:1} \cite{12,12b}, Fig.~\ref{fig:1} makes this point clear. More precisely, the dispersion curve of the classical 2DDS liquid, as generated by our MD simulations, falls in the narrow band of dispersion curves for the 2D bosonic dipole system: the band is bounded from above by the Feynman Ansatz (Eq.~\eqref{eq:1}) and from below by the CBF calculated dispersion.

The remarkable likeness of the strongly coupled bosonic 2DDS dispersion to the excitation spectrum in the superfluid phase of $^4$He \cite{16} has been noted by Kalman {\it et al.} \cite{19}. The most apparent similarity is in the formation of the ``roton minimum'', preceded by the ``maxon'' maximum (see Fig.~\ref{fig:1}). Contrasting, however, with liquid $^4$He, the unique feature of the 2DDS resides in the simplicity of the purely repulsive inter-particle interaction and the possibility of tuning the coupling strength by changing system parameters such as density and temperature.  This makes it possible to analyze the longitudinal collective mode dispersion and its evolution as a function of the $\Gamma_{D}$ and $r_{D}$ coupling parameters defined above and to relate them to the experimentally observed phonon/maxon/roton dispersions in $^4$He. MD simulation studies to this effect have been recently carried out by us to demonstrate that the emergence of the roton minimum in bosonic superfluids is, in fact, a consequence of strong particle correlations and, as such, is basically a classical effect \cite{19}.  This finding gives further impetus to the present  more detailed analysis of the collective mode structure of the 2DDS.

It is in the light of these observations that we have undertaken a detailed combined analytic/computer simulation study of the collective mode behavior of the classical 2DDS liquid and solid phases, whose results are presented in this paper. 

The theoretical calculation of the collective mode dispersion in the liquid phase is to be carried out using the well-tested quasi-localized charge approximation (QLCA) \cite{20,21} with the input of MD-generated pair distribution function $g(r)$ data.  The observation that serves as the basis for the QLCA is that the dominating feature of the physical state of a classical dipolar liquid with coupling parameter $\Gamma_{D} >>1$ is the quasi-localization of the point dipoles.  The ensuing model closely resembles a disordered solid where the dipoles occupy randomly located sites and undergo small-amplitude oscillations about them.  However, the site positions also change and a continuous rearrangement of the underlying quasi-equilibrium configuration takes place.  Inherent in the model theory is the assumption that the two time scales are well separated, and that it is sufficient to consider the time average (converted into ensemble average) of the drifting quasi-equilibrium configuration. The calculation of the lattice phonons in the solid phase is to be carried out using the well-known harmonic approximation.   

MD simulations have been pursued along the lines used for simulating other strongly interacting classical systems \cite{22,SCCS08}. The topical review article \cite{22} provides an in-depth description of the computational methodology followed in the present work.   

In the present work, we study the collective excitations in the 2DDS liquid and solid over a wide range of wave numbers.  In the liquid phase, both the longitudinal and transverse modes have been identified.  In the solid phase, the two modes persist, but the system is anisotropic and the polarizations become wave vector- and angle-dependent. In our earlier work \cite{23}, we have already analyzed the $ka\to 0$ longitudinal acoustic behavior. Here we go further:  Our analysis extends beyond the $ka>2$ domain where a well-defined maximum of the dispersion curve develops followed by a deep minimum.  These extrema are analogous to what has become known as the ``maxon'' and ``roton'' excitations in the mode spectrum of liquid helium.  For this reason, we will use the same terminology throughout this paper.  Our recent MD study of collective modes in the 2DDS solid  \cite{19} indicate the existence of remarkable roton+maxon, maxon+maxon, roton+roton combination frequencies at wave numbers in the vicinities of the maxon and the roton, again, in analogy with what has been theoretically predicted \cite{16,24,24b,24c,24d,24e} in relation to liquid $^4$He spectra. In the present work, we continue this line of investigation for the solid and for the liquid phase as well.  In relation to the phonon spectra we are able to make contact with the elastic theory of solids and identify the principal elastic constants.

The organization of the paper is as follows: In Sec. II, we analyze the collective mode dispersion.  We generate longitudinal ($L$) and transverse ($T$) spectra from our MD simulations of the 2DDS liquid and solid phases.  The ensuing MD dispersion curves, constructed from the peaks of the spectra, serve as standards for comparison with the theoretical QLCA $L$ and $T$ oscillation frequencies displayed in the same section. In Sec. III, we consider the solid phase and apply the harmonic approximation (HA) to the calculation of the longitudinal and transverse acoustic phase velocities, (from which we infer the elastic constants of the crystal), and the full phonon dispersions for the 2DDS. Conclusions are drawn in Sec. IV.

\section{Classical Dipole Liquid: Combined MD/QLCA Analysis }

We turn now to the analysis of the longitudinal and transverse collective modes in the classical 2DDS liquid.  This analysis has been carried out using the QLCA theory combined with MD simulations. 

The successful application of the QLCA to the calculation of collective mode dispersion in a variety of strongly coupled charged particle systems has been well documented over the past two decades.  Its recent application to the 2DDS liquid has resulted in an accurate description of the longitudinal collective mode dispersion in the acoustic domain, as borne out by tabulated comparisons with MD and thermodynamic sound speed data \cite{23}. 

Our program of MD simulation of the dynamics of the classical 2DDS liquid has been carried out along the lines used for simulating other strongly interacting classical systems \cite{22}. The present MD simulation involves 4200 particles; information about the collective modes and their dispersion is obtained from the Fourier transform of the correlation spectra of the microscopic density  
\begin{equation}
n_{k} (t)=\sum _{j}\exp [ikx_{j} (t)],
\end{equation}
yielding the dynamical structure function 
\begin{equation}
S(k,\omega )=(1/2\pi N)\mathop{\lim }\limits_{\Delta \to \infty } (1/\Delta T)\left|n_{k} (\omega )\right|^{2},
\end{equation}
where $\Delta T$ is the duration of the data recording period.  Similarly, the spectra of the longitudinal and transverse current fluctuations, $L(k,\omega)$ and $T(k,\omega)$, can be obtained from Fourier analysis of the microscopic currents \begin{equation}
\lambda_{k} (t)=\sum _{j}v_{jx} (t)\exp [ikx_{j} (t)]
\end{equation}
and 
\begin{equation}
\tau_{k} (t)=\sum_{j}v_{jy} (t)\exp [ikx_{j} (t)];
\end{equation}
we assume that $k$ is directed along the $x$ axis (the system is isotropic). The collective modes are identified as peaks in the fluctuation spectra. The widths of the peaks provide additional information about the lifetimes of the excitations:  narrow peaks correspond to longer lifetimes, while broad features indicate short-lived excitations.

Before proceeding with the analysis, a recap of our recent findings \cite{18,23} for the $ka\to 0$ regime is in order.

The crucial observation concerning the collective behavior of the 2DDS is that average the Hartree field does not exist, since
\begin{equation}
\langle\phi _{D} (r)\rangle_{H} =n\int d^{2} r \phi _{D} (r)
\end{equation}
is unbounded; therefore, the Fourier transform of the dipole potential $\phi _{D} (r)=\mu ^{2} /r^{3} $ does not exist, implying that the 2DDS can have no RPA limit. Comparing this situation with the case of a 2D Coulomb system where either the routine observation
\begin{equation}
\omega^2(k) \propto k^2\phi_{\rm COUL}(k),~~~\phi_{\rm COUL}(k)\propto 1/k,
\end{equation}
or a simple scaling argument for the integral
\begin{equation}
\phi({\bf k}) = \int{\rm d}^2r~\phi(r) \exp(i{\bf k}\cdot{\bf r})
\end{equation}
provide the correct
\begin{equation}
\omega(k)\propto \sqrt{k}
\end{equation}
RPA plasmon behavior, a similar line of reasoning in the present case would lead, via the (faulty) relation
\begin{equation}
\phi_{D}(k)\propto -k^2\phi_{\rm COUL}(k) \propto k,
\end{equation}
to the incorrect RPA longitudinal collective mode frequency as \cite{8}
\begin{equation}
\omega(k)\propto k^{3/2}.
\end{equation}
The correct approach, as established in Ref. \cite{23}, leads to a correlation controlled long-wavelength acoustic dispersion
\begin{equation}
\omega(k)\propto k.
\end{equation}
of the strongly coupled 2DDS liquid both in the classical and in the zero-temperature quantum domains \cite{18,23}. This behavior is also similar to that of the EHB liquid \cite{25}, with phase velocities virtually identical in the two cases. The precise numerical value of the acoustic velocity is governed by the average potential
\begin{equation}\label{eq:int1}
\langle \phi_D(r) \rangle = n\int {\rm d}^{2}r~\phi_{D}(r) g(r),
\end{equation}
rather than by the Hartree potential; $g(r)$ is the statistics- and coupling-dependent equilibrium pair distribution function. We found that the theoretical QLCA values of the longitudinal acoustic phase velocity, calculated from the Eq.~\eqref{eq:int1} integral agree reasonably well with the MD simulation values over a wide range of classical liquid phase coupling strengths; the disparity between the theoretical acoustic phase velocities and thermodynamic sound speeds, though somewhat larger, is still only about 5.3\%. The longitudinal acoustic velocity decreases with increasing coupling parameter and assumes its lowest value in the solid phase. 

The details of the derivation of the longitudinal and transverse dispersions for arbitrary $k$ values based on QLCA theory culminating in the 2DDS dispersion relation \eqref{eq:3} below are given in \cite{23}.  Here it suffices to take Eq.~\eqref{eq:3}, together with the QLCA dynamical tensor \eqref{eq:4}, as the appropriate starting point for the present study:

\begin{equation} 
\label{eq:3} 
\left\| \omega ^{2} \delta _{\mu \nu } -C_{\mu \nu } ({\bf k})\right\| =0,          
\end{equation} 

\begin{eqnarray} \label{eq:4} 
C_{\mu \nu } ({\bf k})&=&-\frac{n}{m} \int {\rm d}^2r~g(r)[\exp (i{\bf k}\cdot {\bf r})-1] \partial _{\mu } \partial _{\nu } \phi _{D} (r) \\ 
&=&\frac{3n\mu ^{2} }{m} \int {\rm d}^2r~\frac{1}{r^{5} } g(r) [\exp (i{\bf k}\cdot {\bf r})-1]\left[\delta _{\mu \nu } -5\frac{r_{\mu } r_{\nu } }{r^{2} } \right] \nonumber 
\end{eqnarray} 

\begin{figure}[htb]
\includegraphics[width=0.9\columnwidth]{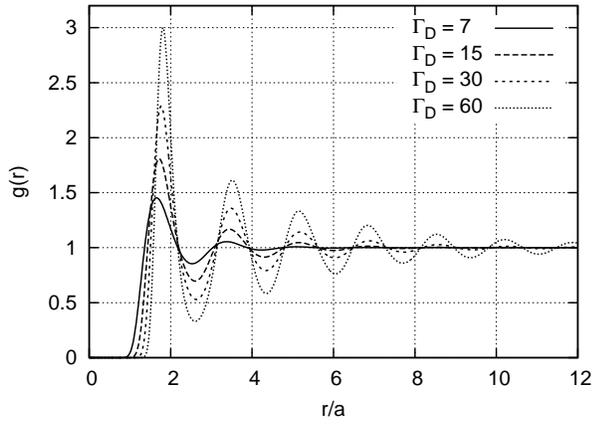}
\caption{\label{fig:2} 
MD pair distribution function for $\Gamma_{D}=7,$ 15, 30, and 60.}
\end{figure}

The longitudinal ($L$) and transverse ($T$) oscillation frequencies are readily calculated to be

\begin{eqnarray}
\label{eq:5} 
\omega _{L}^{2} (k)&=&C_{L} (k) \\ 
&=&\frac{3}{2} \omega _{D}^{2} \int _{0}^{\infty }{\rm d}\bar{r}~\frac{1}{\bar{r}^{4} } g(\bar{r})\left[3-3J_{0} (\bar{k}\bar{r})+5J_{2} (\bar{k}\bar{r})]\right]\nonumber    
\end{eqnarray} 

\begin{eqnarray} 
\label{eq:6} 
\omega _{T}^{2} (k)&=&C_{T} (k) \\
&=&\frac{3}{2} \omega _{D}^{2} \int _{0}^{\infty }{\rm d}\bar{r}~\frac{1}{\bar{r}^{4} } g(\bar{r})\left[3-3J_{0} (\bar{k}\bar{r})-5J_{2} (\bar{k}\bar{r})]\right],\nonumber
\end{eqnarray} 
$\omega _{D}^{2} =2\pi n\mu ^{2} /ma^{3} $, $\bar{r}=r/a$, and $\bar{k}=ka$. Note that the usual interpretation of Eqs.~\eqref{eq:5} and \eqref{eq:6} would require splitting $g(r)$ as
\begin{equation} 
g(r)=1+h(r)
\end{equation} 
and identifying the frequency coming from the ``1'' piece in the integral as the RPA, while $h(r)$ as the correlational contribution. The discussion presented above shows the fallacy of this reasoning for the 2DDS. For $\Gamma _{D}$ fixed, dispersion curves can be generated from \eqref{eq:5} and \eqref{eq:6} with the input of $g(r)$ pair distribution function data obtained from our MD computer simulations. Sample data for the latter are displayed in Fig.~\ref{fig:2}. 

The wave number regimes of special interest are (i) the long-wavelength 
($ka \to 0$) acoustic regime, where both longitudinal and transverse acoustic modes develop, (ii) the finite wave number regime spanning the maxon-roton portion of the longitudinal dispersion curve, and (iii) the high-$k$ domain where the dispersion is dominated by single particle excitations.

Addressing first the long-wavelength ($ka\to 0$) regime, Eqs. \eqref{eq:5} and \eqref{eq:6} simplify to

\begin{equation} \label{eq:7} 
\omega _{L}^{2} (k\to 0)=\frac{33}{16} J(\Gamma _{D} )\omega _{D}^{2} \bar{k}^{2} ,        
\end{equation} 

\begin{equation} \label{eq:8} 
\omega _{T}^{2} (k\to 0)=\frac{3}{16} J(\Gamma _{D} )\omega _{D}^{2} \bar{k}^{2} ,        
\end{equation} 

\begin{equation} \label{eq:9} 
J(\Gamma _{D} )=\frac{1}{2} \frac{\langle\phi _{D} (r)\rangle}{\phi _{D} (a)} =\int _{0}^{\infty }d\bar{r}\frac{1}{\bar{r}^{2} } g(\bar{r}) .       
\end{equation} 

\begin{table}[b]
\caption{\label{tab:1}
2D-point dipole liquid:  QLCA ($s_{L,T}^{\rm QLCA} $), MD ($s_{L,T}^{\rm MD} $), and thermodynamic ($s^{\rm TH} $) sound speeds as functions of the classical coupling parameter $\Gamma _{D}$. The entries in columns 2-5 are quoted from \cite{23} for $\Gamma_{D} =10-40$, $\omega_D=\sqrt{2\pi n\mu^2/ma^3}$.}
\begin{ruledtabular}
\begin{tabular}{ccccccc}
 $\Gamma _{D}$ & $J(\Gamma _{D})$ & $s_{L}^{\rm QLCA}$ & $s_{L}^{\rm MD}$ & $s_{L}^{\rm TH}$ & $s_{T}^{\rm QLCA}$ & $s_{T}^{\rm MD}$\\
 &&($a\omega _{D} $)&($a\omega _{D} $)&($a\omega _{D} $)&($a\omega _{D} $)&($a\omega _{D} $)\\
\hline
10 & 0.8847 & 1.351 & 1.312 & 1.282 & 0.4073 & - \\
20 & 0.8504 & 1.324 & 1.276 & 1.257 & 0.3992 & - \\
30 & 0.8370 & 1.314 & 1.246 & 1.247 & 0.3962 & 0.38 \\ 
40 & 0.8295 & 1.308 & 1.258 & 1.242 & 0.3944 & 0.38 \\ 
60 & 0.8208 & 1.301 & 1.256 & 1.234 & 0.3923 & 0.37 \\ 
\end{tabular}
\end{ruledtabular}
\end{table}

The values of $J(\Gamma _{D})$ calculated from \eqref{eq:9} with the input of MD-generated pair distribution function data and the ensuing values of the QLCA acoustic velocities calculated from Eqs. \eqref{eq:7} and \eqref{eq:8} are displayed in Table I along with the corresponding MD acoustic phase velocities and the thermodynamic sound speeds ($s^{\rm TH}$) calculated from the isothermal compressibility using the classical 2DDS equation of state \cite{23}.  Using the MD data as a reference, the discrepancy between longitudinal QLCA and MD sound speeds ranges from 2.89\% at $\Gamma _{D} =10$ to 3.46\% at $\Gamma_{D} =60$.  The entries for the MD transverse acoustic speeds in column 7 are calculated as the differential slope $s=\partial \omega /\partial k$ in the linear regime at $ka \approx 1$. Such data for $\Gamma _{D} \le 20$ are not available because in this weaker coupling regime the shear wave is too heavily damped, similarly to the situation for the 2DOCP \cite{26}.  The discrepancy between the transverse QLCA and MD acoustic velocities ranges from 4.09\% at $\Gamma _{D} =30$  to 5.68\%  at $\Gamma _{D} =60$.

\begin{figure}[htb]
\includegraphics[width=0.8\columnwidth]{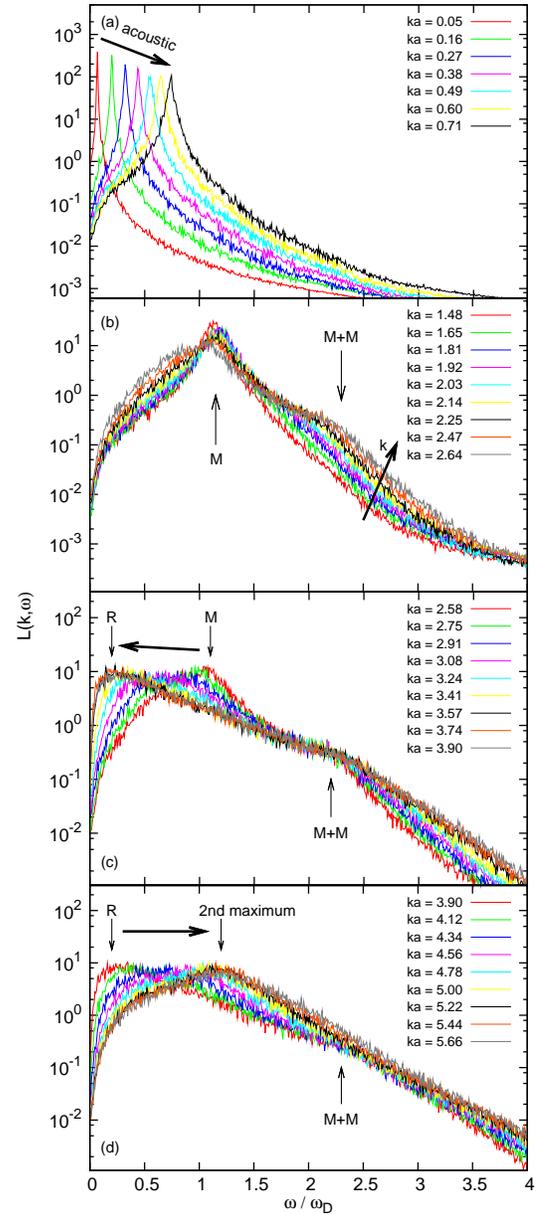}
\caption{\label{fig:3} 
(color online) Longitudinal ($L$) current fluctuation spectra for $\Gamma _{D} =60$;  $a=1/\sqrt{\pi n}$ is the 2D Wigner-Seitz radius and $\omega _{D} =\sqrt{2\pi n\mu ^{2} /ma^{3} } $ is a characteristic dipole oscillation frequency.  The thick arrows point in the direction of increasing $ka$.}
\end{figure}

In Table \ref{tab:1}, note that multiplication by $\sqrt{2\Gamma _{D}}$ converts $a\omega_{D} $ units into $1/\sqrt{\beta m}$ thermal velocity units; $\omega_D=\sqrt{2\pi n\mu^2/ma^3}$ is a characteristic dipole oscillation frequency. In these latter units, the thermodynamic speed decreases from a value of 13.52 at $\Gamma _{D} =60$ to unity at $\Gamma_{D} =0$.  We observe that the MD and QLCA longitudinal phase velocities in Table \ref{tab:1} are somewhat higher than the thermodynamic speeds.  The analysis of the 2DDS crystal in Sec. 3 may shed some light on the origin of this difference.   

\begin{figure}[htb]
\includegraphics[width=0.8\columnwidth]{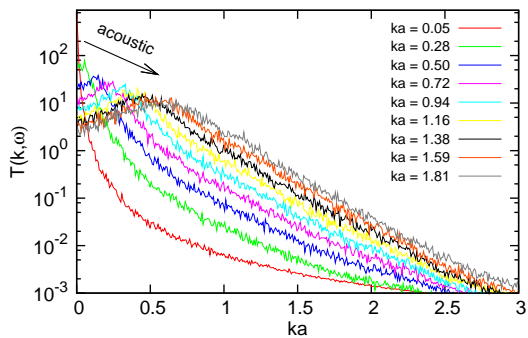}
\caption{\label{fig:4} 
(color online) Transverse ($T$) current fluctuation spectra for $\Gamma _{D} =60$.}
\end{figure}

Turning next to the finite $ka$ domain, we have generated MD spectra for a wide range of $\Gamma_D$ values. Representative $L(k,\omega)$ and $T(k,\omega)$ spectra at $\Gamma_D=60$ are displayed in Figs.~\ref{fig:3} and \ref{fig:4}. In Fig.~\ref{fig:3}  there are four $ka$ domains between $ka = 0.05$ and $ka = 5.66$.  Looking at Fig.~\ref{fig:3}a we observe that the spectral peak shifts to higher frequencies with increasing $ka$ values, reaching the maxon frequency $\omega_M \approx 1.2 \omega_D$ around $ka \approx 2$ (Fig.~\ref{fig:3}b). For $ka$ values in the sub-interval $[1.48, 2.64]$ depicted in Fig.~\ref{fig:3}b, we observe two well-defined clusters of spectral peaks: the lower frequency cluster represents the domain of the maxon; within the higher frequency cluster, the broad peaks indicate the emergence of a faint maxon-maxon (M+M) harmonic. For $ka$ values in the sub-interval $[2.58, 3.90]$ depicted in Fig.~\ref{fig:3}c, the lower frequency peaks shift to the domain of the roton minimum which assumes its lowest frequency value at around $ka \approx 3.7$; we note the persistence of the maxon+maxon harmonic over this entire $ka$ sub-interval.  At the higher $ka$ values depicted in Fig.~\ref{fig:3}d, the cluster of lower frequency spectral peaks indicates the existence of a second maximum at around $ka \approx 5.4$.  

\begin{figure}[htb]
\includegraphics[width=\columnwidth]{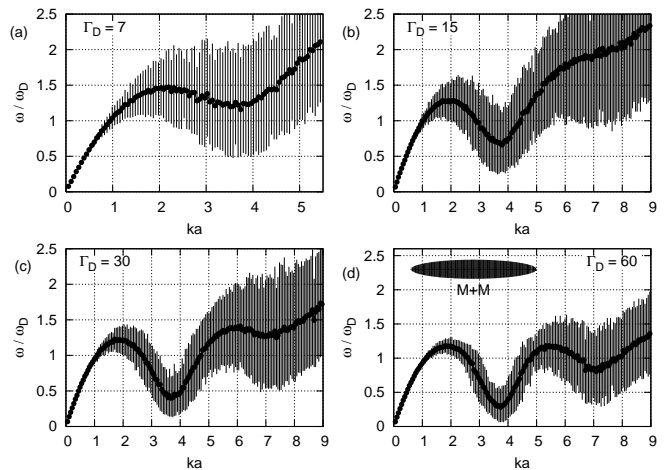}
\caption{\label{fig:5} 
Longitudinal dispersion curves for $\Gamma _{D} =7, 15, 30,$ and 60.  The vertical lines indicate the widths of the spectra.  The shaded region in (d) should be regarded as a ghost-like dispersion depicting the emergence of a faint maxon-maxon (M+M) harmonic.}
\end{figure}

\begin{figure}[htb]
\includegraphics[width=0.9\columnwidth]{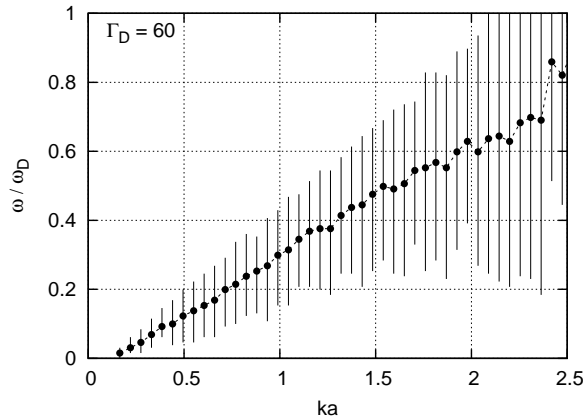}
\caption{\label{fig:6} 
Transverse shear mode dispersion curve for $\Gamma _{D} =60$.  Note the wave number cutoff $k^*a$ as a function of $\Gamma _{D}$ marking the $ka$ value where $\omega (ka)=0$.}
\end{figure}

The resulting MD longitudinal and transverse dispersion curves obtained from the peaks of the $L$ and $T$ spectra, respectively, are displayed in Figs.~\ref{fig:5} and \ref{fig:6}. 

Addressing first the longitudinal mode dispersion in Fig.~\ref{fig:5}, we observe the progressive deepening of the roton minimum with increasing $\Gamma_{D}$.  The broadening of the spectral peaks becomes more and more pronounced with decreasing coupling strength; that is, the lifetime of the collective mode decreases with increasing temperature.  The simulations indicate that, for wave numbers $0 \le ka < 2.5$, spanning the entire acoustic domain and extending somewhat beyond the maxon peak, the longitudinal mode is fairly robust.  At higher wave numbers, the mode is strongly damped, but can still be viable over the approximate domain $2.5<ka<5$ so long as the 2DDS liquid is in the $\Gamma_{D} \ge 30$ strong coupling regime. At $\Gamma_{D} =60$, we observe from Fig.~\ref{fig:5} that the location of the roton minimum in the classical 2DDS liquid is close to the location of its counterpart in the 2D bosonic dipole liquid at zero temperature \cite{12,12b}.
In the Conlusions we further discuss the relevance of the observed roton-maxon behavior to the dispersion characteristics of the 2D bosonic dipole liquid. 

In the high-$ka$ domain, we observe that the longitudinal dispersion eventually becomes dominated by the Bohm-Gross (BG) oscillation frequency
\begin{equation} \label{eq:Bohm}
\omega(k)=\frac32\frac{k}{\sqrt{\beta m}}=\frac32 \omega_D\frac{ka}{\sqrt{2\Gamma_D}}
\end{equation}
characteristic of single-particle excitations in the classical 2DDS, and in strict compliance with the third-frequency-moment sum rule for dipole systems \cite{KGGJKu}. The switchover $ka$ value for the single particle beavior occurs between $ka= 5$ (for $\Gamma_D=15$) and $ka=8.5$ (for $\Gamma_D=60$). Again, making contact with the 2D bosonic dipole liquid, it is of some interest to compare this $\omega(k\to \infty)\propto k$ asymptotic behavior with the one predicted by the Bogolyubov theory for the quasicondensate 2DDS \cite{18,22}. In this latter the high-$ka$ behavior sets on around similar $k$-values \cite{18} and is also dominated by single-particle excitations, which, however, originate, in sharp contrast to the classical system, from the zero momentum ground state. Hence the difference in the asymptote, which in the zero temperature limit is $\omega(k\to \infty)\propto \hbar k^2/2m$.

The shaded region in Fig.~\ref{fig:5}(d) is seen as the emergence of the faint maxon-maxon (M+M) harmonic detected in the Fig.~\ref{fig:3} spectra. More will be said about this harmonic, along with other combination frequencies that emerge only in the 2DDS lattice, in Sec. III.   

We next address the transverse shear mode dispersion displayed in Fig.~\ref{fig:6}.  Similarly to the 2D one-component plasma (2DOCP) \cite{26} and to complex plasmas \cite{27,28,29}, shear waves in the strongly coupled 2DDS liquid are strongly damped, even at $\Gamma _{D} =60$, as evidenced by the sizeable line widths shown in Fig.~\ref{fig:6}; for $\Gamma _{D} <30$, the MD simulations indicate that the shear waves are too heavily damped to be viable.  In the coupling regime where they are viable, Fig.~\ref{fig:6} and Fig.~\ref{fig:9} below show that, similarly to the 2DOCP \cite{26} and to 2D Yukawa plasmas (2DYP)  \cite{27,28,29}, 2DDS liquid-phase shear waves cease to exist below a critical finite $\Gamma _{D}$-dependent wave number cutoff, $k^*a$, marking the $ka$ value where $\omega (ka)=0$. As expected, $k^*a\to 0$ as $\Gamma _{D} \to \Gamma _{D}^{\rm SOLID}$ from below.

\begin{figure}[htb]
\includegraphics[width=0.9\columnwidth]{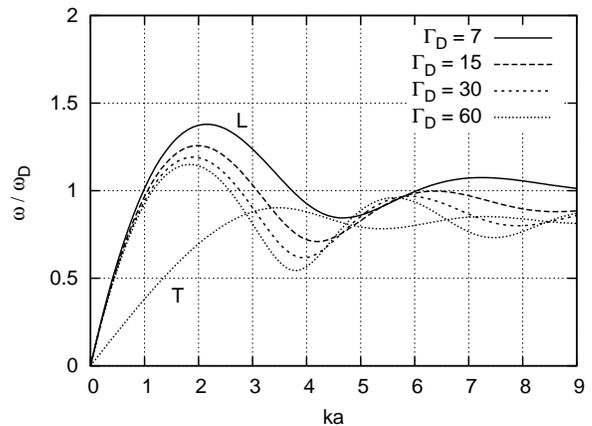}
\caption{\label{fig:7} 
QLCA Longitudinal ($L$) and transverse ($T$) dispersion curves for $\Gamma _{D} =7,$ 15, 30, and 60.  Note the evolution of the roton minimum with increasing$\Gamma _{D}$.}
\end{figure}

\begin{figure}[htb]
\includegraphics[width=\columnwidth]{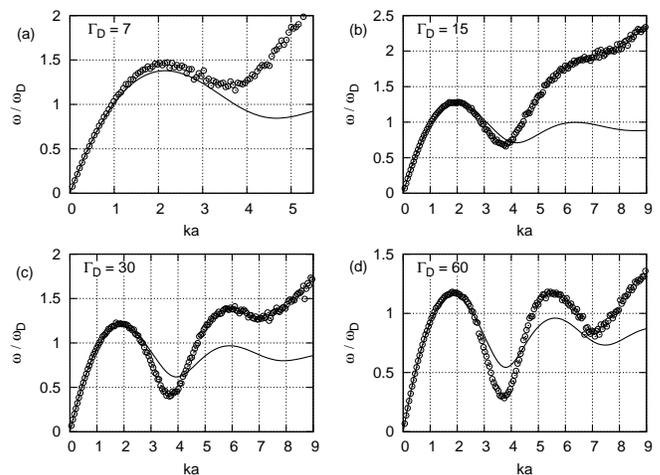}
\caption{\label{fig:8} 
MD and QLCA longitudinal dispersion curves for $\Gamma_{D} =7,$ 15, 30, and 60.}
\end{figure}

Turning now to the QLCA description of the mode dispersion at finite wavenumbers, the straightforward calculation of the QLCA oscillation frequencies \eqref{eq:5} and \eqref{eq:6}, with the input of MD-generated $g(r)$ data, results in the longitudinal and transverse dispersion curves displayed in Fig.~\ref{fig:7} for $\Gamma _{D} =7$, 15, 30, and 60. Figs.~\ref{fig:8} and \ref{fig:9} provide a comparison between the QLCA dispersion curves and the MD data.  

Figure~\ref{fig:8} shows very good quantitative agreement between theory and simulation up to $ka \sim 2.5$. As expected, the $ka$ range of agreement increases with increasing coupling, since the QLCA theory is, after all, premised to be a strong coupling theory. For increasing $ka$-values, we can observe the evolution of the roton minimum with increasing $\Gamma_{D}$. According to the QLCA description, the position of the roton minimum shifts to lower $ka$ values and deepens more and more with increasing coupling strength. The MD data, however, indicate that the $ka \sim 3.7$ position of the roton minimum remains more or less the same as it progressively deepens with increasing coupling strength.  While the QLCA does capture the qualitative features of the roton portion of the 2DDS dispersion curve for $\Gamma_{D} \ge 30$, quantitative agreement between theory and simulation for $ka >3$ is less satisfactory, even in this high coupling regime of the liquid phase.   

\begin{figure}[htb]
\includegraphics[width=0.9\columnwidth]{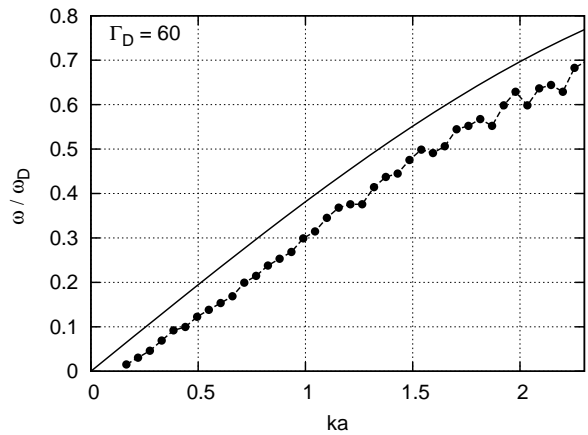}
\caption{\label{fig:9} 
MD and QLCA transverse dispersion curves for $\Gamma_{D} =60$.}
\end{figure}

Turning now to the transverse shear mode, Fig.~\ref{fig:9} shows only fair agreement between the QLCA and MD dispersion curves.  Similarly to what has been reported for 2D Yukawa liquid plasmas \cite{28}, the MD data, anchored by the ($k^*a$, $\omega =0$) point, lie to the right of the QLCA dispersion curve and run parallel to it in the linear regime. This discrepancy is due to the QLCA's inability to account for diffusional and other damping effects that preclude the existence of long-wavelength shear waves in the liquid phase \cite{30}.  In the case of 2D Yukawa liquids \cite{28}, the discrepancy was removed by introducing a phenomenological diffusional damping time into the QLCA formalism.  We expect that a similar such modification in the 2DDS QLCA formalism would bring about the same outcome.

\section{Dipole Lattice: Harmonic Approximation and MD Simulations}

According to our MD simulation the 2DDS liquid freezes at about $\Gamma_D \sim 70$. In this Section we turn to the analysis of the collective modes (phonon dispersion) of the lattice phase. As in all systems with an isotropic central force interaction, the 2DDS crystallizes in a triangular (hexagonal) lattice. We calculate the phonon dispersion in the harmonic approximation (HA). The starting point is the dispersion relation \eqref{eq:1} with the liquid-phase dynamical tensor \eqref{eq:4} replaced by its HA counterpart \cite{23}

\begin{eqnarray} \label{eq:10} 
C_{\mu \nu } ({\bf k})&=&-\frac{1}{m} \sum _{i}[\exp (i{\bf k}\cdot {\bf r}_{i} )-1]\partial _{\mu } \partial _{\nu } \phi _{D} (r_{i} ) \\
&=&\frac{3\mu ^{2} }{m} \sum _{i}[\exp (i{\bf k}\cdot {\bf r}_{i} )-1] \frac{1}{r_{i}^{5} } \left[\delta _{\mu \nu } -5\frac{r_{i\mu } r_{i\nu } }{r_{i}^{2} } \right], \nonumber
\end{eqnarray} 
where ${\bf k}\cdot {\bf r}_{i} =kr_{i} \cos(\theta_{i}-\varphi); \theta_{i}$ is the angle of the ${\bf r}_i$ vector and $\varphi$ is the propagation angle measured from the axis pointing toward the nearest neighbor.

At long wavelengths, the hexagonal lattice is isotropic so that the direction of \textbf{k} is arbitrary.  So for convenience, one can choose \textbf{k} to be along one of the two crystallographic axes, say, $\varphi_ =0^{0}$. In this case Eq.~\eqref{eq:10} then simplifies to  

\begin{equation} 
\label{eq:11} 
C_{\mu \nu } (k\to 0)=-\frac{3\mu ^{2} k^{2} }{m} \sum _{i}\cos ^{2} \theta _{i}  \frac{1}{r_{i}^{3} } \left[\delta _{\mu \nu } -5\frac{r_{i\mu } r_{i\nu } }{r_{i}^{2} } \right].
\end{equation} 
Moreover, for the isotropic system, the dynamical tensor \eqref{eq:11} is diagonal, implying that the eigenvectors are parallel and perpendicular to \textbf{k}, as in the liquid.  The purely longitudinal and transverse oscillation frequencies then follow from the dispersion relation \eqref{eq:3} and \eqref{eq:11}:

\begin{eqnarray} \label{eq:12} 
\omega _{L}^{2} (k\to 0)&=&C_{L} (k\to 0)= \\
&&-\frac{3\mu ^{2} k^{2} }{m} \sum _{i}\cos ^{2} \theta _{i}  \frac{1}{r_{i}^{3} } \left[1-5\cos ^{2} \theta _{i} \right], \nonumber 
\end{eqnarray} 

\begin{eqnarray} \label{eq:13} 
\omega _{T}^{2} (k\to 0)&=&C_{T} (k\to 0)= \\
&&-\frac{3\mu ^{2} k^{2} }{m} \sum _{i}\cos ^{2} \theta _{i}  \frac{1}{r_{i}^{3} } \left[1-5\sin ^{2} \theta _{i} \right]. \nonumber 
\end{eqnarray} 

The hexagonal lattice can be decomposed into a sequence of concentric hexagons tilted away from the crystallographic axes and of increasing size. Thus the lattice sum \eqref{eq:12} can be decomposed into an $r_i$-dependent part and an angular part; this latter can be summed over the vertices of each of the hexagons. This summations is facilitated by observing that for any arbitrary angle $\alpha$, the sum $\sum_{n=0}^{5} \cos^{2} (\alpha +n\pi /3)=1/2 $ is independent of $\alpha$. It therefore suffices to sum the $\theta _{i} $ portion of the summands in \eqref{eq:12} and \eqref{eq:13} around a single hexagon and divide by six. This results in

\begin{equation} \label{eq:14} 
\omega _{L}^{2} (k\to 0)=C_{L} (k\to 0)=\frac{33}{32} M\omega _{D}^{2} a^{2} k^{2} ,       
\end{equation} 

\begin{equation} \label{eq:15} 
\omega _{T}^{2} (k\to 0)=C_{T} (k\to 0)=\frac{3}{32} M\omega _{D}^{2} a^{2} k^{2},       
\end{equation} 
where $M=\sum _{i}1/\bar{r}_{i}^{3}$ is the lattice sum over the triangular lattice with $\bar{r}_{i} \equiv r_{i} /a$.  The longitudinal oscillation frequency \eqref{eq:14} has been reported in \cite{23} and is displayed here alongside the new transverse result \eqref{eq:15} for comparison.  In effect, the lattice sum $M/2$ replaces its liquid-phase counterpart integral $J(\Gamma _{D})$ in Eqs. \eqref{eq:7} and \eqref{eq:8}.  The value of $M$ has been calculated by a number of workers \cite{31,32,33,34} with slightly different results; the most recent semi-analytical calculation due to Rozenbaum \cite{34} is quoted here as $M=1.642$.  Our own lattice sum computation for the 2DDS crystal involving $1.9\times 10^{9}$ particles provides $M=1.597$ \cite{19}. From Eqs. \eqref{eq:14} and \eqref{eq:15}, the corresponding $L$ and $T$ sound speeds are

\begin{equation} \label{eq:16} 
s_{L} =1.283\omega _{D} a,          
\end{equation} 

\begin{equation} \label{eq:17} 
s_{T} =0.387\omega _{D} a.          
\end{equation} 

As expected, these values are slightly lower than their respective $\Gamma _{D} =60$ liquid phase counterpart entries in Table \ref{tab:1}.  Note the agreement between \eqref{eq:16} and the measured value $1.2836 a\omega_{D}$. The thermodynamic sound speed $s^{\rm TH} $, though not physically meaningful for a lattice, is still of interest from the point of view of providing an estimate of the liquid phase thermodynamic sound speed in the large-$\Gamma_{D}$ limit.  Table~\ref{tab:1} indicates that the thermodynamic sound speed decreases monotonically with increasing $\Gamma_{D}$.  One therefore expects that $s^\text{TH}$ assumes its minimum value when $\Gamma_{D} \to \infty$.  

For an isotropic elastic medium the long wavelength ($k \to 0$) behavior of the phonons can be described in terms of longitudinal and transverse elastic waves, rather than in terms of the thermodynamic sound speed, $s^\text{TH}$, with phase velocities expressed in terms of the  elastic constants $K$ and $G$. $K=-V(\partial P/\partial V)$ is the bulk modulus, $G$ is the shear modulus. In 2D the corresponding velocities are
\begin{eqnarray} \label{eq:18}
s^\text{TH} &=& \sqrt{\frac{K}{\rho}} \nonumber \\
s_L &=& \sqrt{\frac{K+G}{\rho } } \nonumber \\
s_T &=& \sqrt{\frac{G}{\rho } }  
\end{eqnarray}
with $\rho =mn$ being the mass density of the solid. It is important to note the marked difference between the thermodynamic sound veloicity $s^{\rm TH}$ and the elasic phase velocities $s_{L}$ and $s_{T}$ as enunciated by \eqref{eq:18}. There is a small, but discernible, difference between the $s^{\rm TH}$ and $s_{L}$ and $s_{T}$,
\begin{equation} \label{eq:19} 
s^{\rm TH} =\sqrt{s_{L}^{2} -s_{T}^{2} }  
\end{equation} 
readily follows from \eqref{eq:18}.  From \eqref{eq:16} and \eqref{eq:17}, the ratio $s_{T}^{2} /s_{L}^{2} =1/11$ then provides
\begin{equation} \label{eq:20} 
s^{\rm TH} =s_{L} /1.049=1.223a\omega_{D}  
\end{equation} 
for the 2DDS hexagonal lattice.  This value is in keeping with the trend in Table \ref{tab:1}.    

\begin{figure}[htb]
\includegraphics[width=\columnwidth]{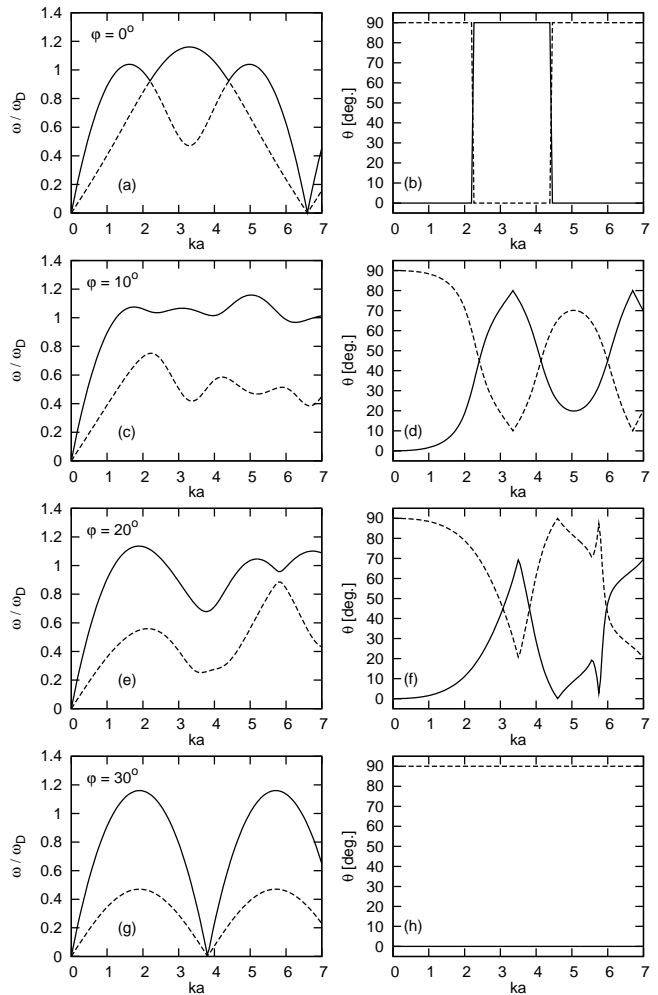}
\caption{\label{fig:10} 
Electric dipole hexagonal lattice eigenmodes and corresponding mode polarizations for angular directions (with respect to the nearest neighbor direction) $\varphi =0-30^{0} $ in $10^{0}$ steps. The measured sound velocity $s_{L} =1.284a\omega_{D}$.}
\end{figure}

We address next the lattice phonon dispersion at finite wavenumbers obtained by solving the dispersion relation \eqref{eq:3} with the input of the dynamical tensor \eqref{eq:10}. The results very closely resemble the dispersion of phonon spectrum of another known 2D system, the 2D Yukawa triangular (hexagonal) lattice \cite{22}. In particular, the value $1/11$ of the sound speed ratio $s_T^2 /s_L^2$ turns out to be the same as the sound speed ratio for the 2D Yukawa crystal \cite{35} with a screening parameter $\kappa a=1.05$. Phonon dispersion curves displayed in Fig. \ref{fig:10} for four propagation angles: $\varphi =0^{0} ,10^{0} ,20^{0} ,30^{0}$ ($\varphi =0^{0}$ and $\varphi =30^{0}$ are the crystallographic axes). The angle $\Theta$, indicated in the right panels of Fig. \ref{fig:9}, is the polarization angle measured with respect to the propagation vector \textbf{k}; the mode polarizations are purely longitudinal ($\Theta =0^{0}$) or purely transverse ($\Theta =90^{0}$) for propagation along the $\varphi =0^{0}$ and $\varphi =30^{0}$ crystallographic axes only. Otherwise, the polarizations are mixed as shown in the $\varphi =10^{0}$ and $\varphi =20^{0}$ right panels of Fig.~\ref{fig:9}. The dispersion curves are periodic in $k$; the period is the reciprocal lattice constant only for propagation along a crystallographic axis; for intermediate angles, the much longer period is given by formulas (45) and (46) in \cite{22}.

For a macroscopically disordered lattice one can view the system as an aggregate of locally ordered domains with randomly distributed crystallographic axes.  One may also contemplate this as an alternate model for the strongly coupled liquid. We can then seek the similarity to the liquid-phase dispersion through a suitably angle-averaged dispersion of the lattice. The procedure, which parallels that of Ref.~\cite{22}, consists in projecting out the longitudinal and transverse components of the lattice eigenmodes and comparing their respective angular averages with the longitudinal and transverse modes in the liquid phase.  The calculation of the angle-averaged longitudinal and transverse lattice dispersions proceeds according to the following prescription:  First, we compute the lattice eigenmodes for \textbf{k} vectors with angular directions between $\varphi =0^{0}$ and $\varphi =30^{0} $ in one degree steps.  We next perform the longitudinal and transverse projections based on the normal mode data with 

\begin{eqnarray}
\omega_L^2&=&\left({\bf \hat{k} \cdot \hat{e}_1}\right)^2\omega_1^2 +
\left({\bf \hat{k}\cdot \hat{e}_2}\right)^2\omega_2^2 \\
\omega_T^2&=&\left[1-\left({\bf \hat{k} \cdot
\hat{e}_1}\right)^2\right]\omega_1^2 + \left[1-\left({\bf \hat{k}\cdot
\hat{e}_2}\right)^2\right]\omega_2^2,\nonumber 
\end{eqnarray}
where ${\bf \hat{k}}$, ${\bf \hat{e}_1}$ and ${\bf \hat{e}_2}$ are
unit vectors parallel to the wave vector (${\bf k}$) and normal-mode
eigenvectors (${\bf e_1}$ and ${\bf e_2}$), respectively; $\omega_{1} $ and $\omega_{2}$ are the normal-mode frequencies \cite{36}.  We can then simply average the frequency values belonging to equal $k$ values for the longitudinal and transverse dispersions separately.

\begin{figure}[htb]
\includegraphics[width=0.9\columnwidth]{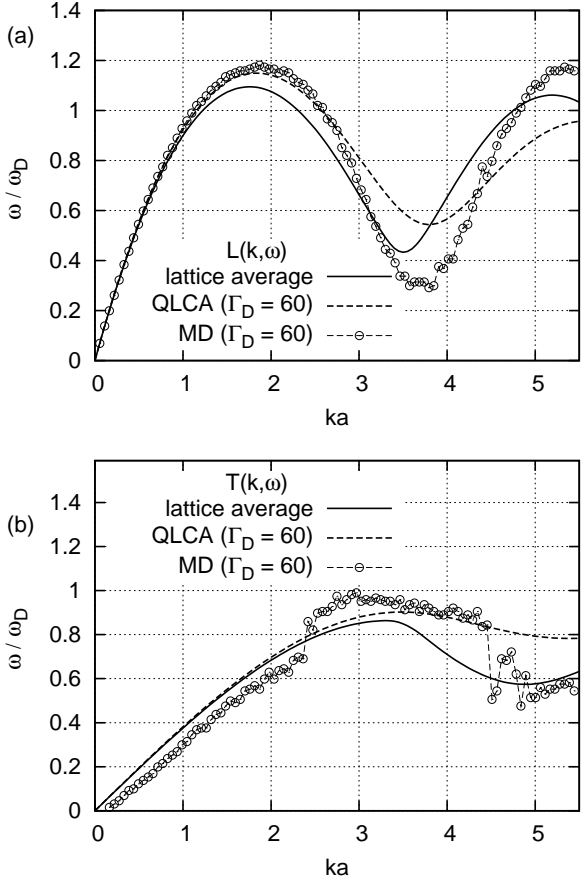}
\caption{\label{fig:11} 
2DDS lattice dispersion.  Angle-averaged longitudinal and transverse lattice modes shown with liquid phase MD data and the QLCA dispersion curves at $\Gamma_{D} =60$.}
\end{figure}

The results are displayed in Fig.~\ref{fig:11}. Addressing first the agreement between the two approaches and the MD we see that the QLCA provides a superior descrition for $0 \leq ka \leq 2.8$ and somewhat less satisfactory quantitative agreement thereafter. In contrast the lattice average dispersion, while it shows excellent agreement with the MD data over the narrower interval $0 \leq ka \leq 1$, compares somewhat less favorably with the MD data than the QLCA thereafter, but becomes superior to the QLCA in the neighborhood of the high-$k$ roton region. As to the transverse collective mode dispersion shown in Fig.~\ref{fig:11}b, we observe the same trend. The discrepancy between the QLCA dispersion and the MD data was discussed earlier in connection with the finite $ka$ cutoff.

\begin{figure}[t]
\includegraphics[width=\columnwidth]{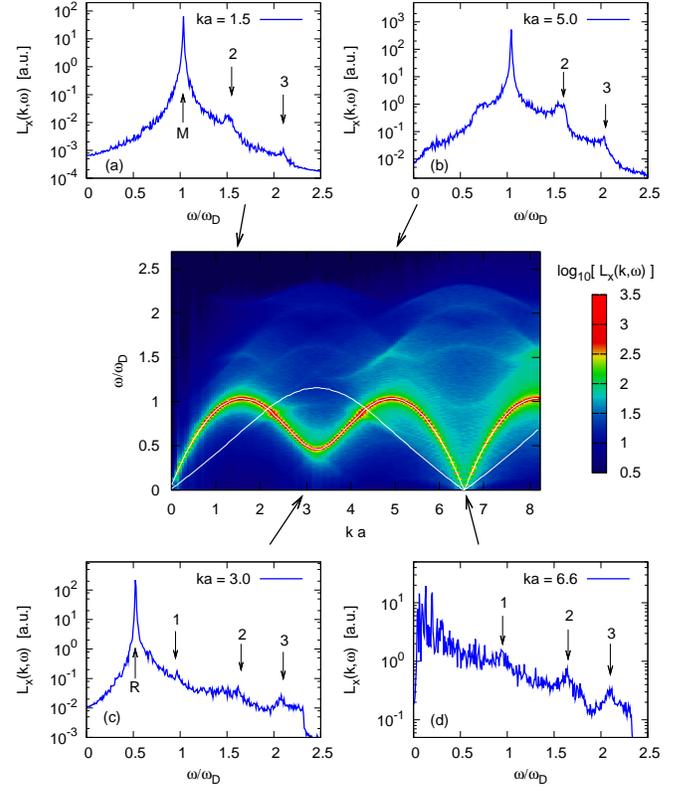}
\caption{\label{fig:12} 
(color online) Illustration of the appearance of combination frequencies in the longitudinal current-current correlation functions $L(k=k_x,\omega)$, measured along the direction of the nearest neighbor in a hexagonal lattice simulated at $\Gamma_D = 1000$. The center color map of the $L(k=k_x,\omega)$ fluctuation spectra shows the strong primary dispersion and the ghosts of the combination frequencies, thin white lines are the dispersion curves from both $L(k=k_x,\omega)$ and $T(k=k_x,\omega)$. The four small panels show vertical cross-sections taken for the selected group of wave numbers, where the appearance of the combination frequencies is the most manifest.}
\end{figure}

\begin{figure}[t]
\includegraphics[width=\columnwidth]{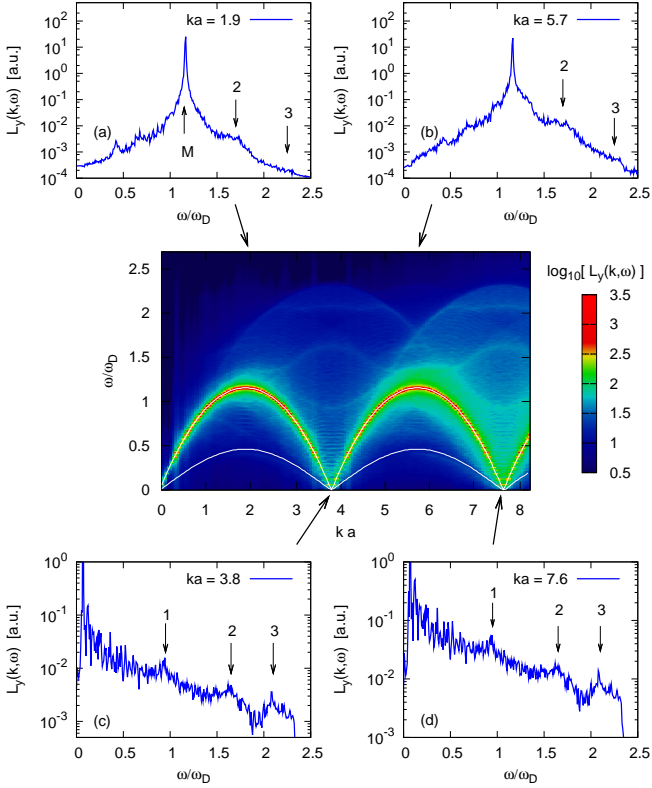}
\caption{\label{fig:13}
(color online) Same as Fig.~\ref{fig:12} for wave-numbers perpendicular to the nearest neighbor direction, $L(k=k_y,\omega)$.}
\end{figure}

In Figs.~\ref{fig:12} and \ref{fig:13}, we have displayed a series of $L(k=k_x,\omega)$ and $L(k=k_y,\omega)$ spectra generated over a wide range of $ka$ values at $\Gamma_{D} = 1000$.  Our findings reaffirm the observations from our previous simulations \cite{19} carried out at $\Gamma_D = 500$ and over a narrower range of wave numbers:  The accumulation of weight in the vicinity of three combination frequencies,   labeled as ``1'', ``2'' and ``3'' is clearly visible.  The emergence of harmonics appears to be a general feature of Yukawa, Coulomb, and other types of similar interaction \cite{37} as well, although the quantitative relationships between the amplitudes of the different combination frequencies should be sensitive to the actual interaction potential.   

The identification of the ``1'', ``2'', ``3'' peaks with the roton-roton (R+R), maxon-roton (M+R and perhaps M-R), maxon-maxon (M+M) combinations, respectively, is tempting, but the full understanding of this process requires more study. It should be realized that due to the anisotropy of the medium, the formation of combination frequencies in a lattice is a much more intricate matter than in an isotropic liquid. First, since the positions of the minima and maxima of the dispersion curves are angle dependent, the notions of roton frequency and maxon frequency become ill-defined and the strengths and positions of the combination frequencies become also angle dependent; this is well illustrated by the comparison of Figs.~\ref{fig:12} and \ref{fig:13} where modes propagating along the two different crystallographic axes are compared. Second, most of the weight should now originate from the vicinity of the local extrema of the two-dimensional dispersion surface, rather than from the extrema of the selected  dispersion curves; but the nature and locations of these former have not yet been explored. Third, the appropriate matching of the $k$-vectors may become a delicate issue, and one would expect that directions in which the highest amplitude harmonics appear may be quite different from the directions associated with their parent frequencies. In fact, as pointed out by \cite{24d}, it is the very close proximity of the $k=0$ region where the largest possible number of combination \textbf{k}-vectors could end up, thus leading to the expectation that high-amplitude combination frequencies would appear in this domain. It is therefore remarkable that our data do not indicate the presence of any weight in the $k=0$ region: whether this is due to the limitations of the simulation protocol or a real effect, still remains to be understood.

\section{Conclusions}

In this paper we have carried out a combined analytic/molecular dynamics (MD) study of the dynamics and collective mode dispersion of a classical two-dimensional dipole system (2DDS) in its strongly coupled liquid and solid phases. The analytical methodology is based on the quasi-localized charge approximation (QLCA) for the description of the liquid-phase dispersion, and on the companion harmonic approximation (HA) for the description of the lattice phonons. As we stated in the Introduction to this paper, we had two objectives in mind:  The first objective was to understand how the dynamics of classical many-body system with a $1/r^3$ interaction potential are related to the dynamics of similar systems, such as the 2D OCP and the 2D Yukawa plasma. The second objective was to relate the collective mode behavior of the classical 2DDS to that of its quantum counterpart, the 2D bosonic dipole system.  

As to the first objective, in contrast to the 2D OCP which features an $\omega(k\to 0)\propto \sqrt{k}$ longitudinal mode, the 2DDS exhibits an $\omega(k\to 0)\propto k$ acoustic excitation. In this sense, the behavior of the 2DDS is similar to that of the 2D Yukawa system. The 2DDS, however, is unique in that the longitudinal acoustic mode is wholly maintained by strong short-range correlations. 

Similarly to the 2DOCP \cite{18} and complex plasmas \cite{20,21,22}, our MD simulations reveal that transverse shear waves propagating in the strongly coupled 2DDS liquid are strongly damped, especially for coupling strengths $\Gamma_D<60$. Moreover, the waves cease to exist below a critical wavenumber cutoff, marking the point where the excitation frequency becomes identically zero. As expected, this critical wave number decreases with increasing coupling strength, tending to zero when the 2DDS liquid freezes.  

Using the harmonic approximation, we have calculated the longitudinal and transverse phonon dispersions of the crystalline solid phase. The morphology of the lattice phonons closely emulates that of the 2D Yukawa lattice; this is not surprising since the morphology appears to be dictated primarily by a common hexagonal lattice structure rather than by the nature of the interaction. For the macroscopically disordered lattice, where the system can be viewed as an aggregate of locally ordered domains whose symmetry axes are randomly distributed, we have calculated the angle-averaged lattice dispersion curves. One may also contemplate this approach as a model alternate to the QLCA for the strongly coupled liquid. The comparison of the two of them with the MD results show reasonable agreement up to high $ka$ values: the QLCA seems to perform better in the  lower, while the averaged lattice model in the higher $ka$ regions.

Addressing the second objective, we note that the main interest in this regard lies in the understanding of the formation of the so-called maxon-roton structure of the dispersion in the 2D bosonic dipole liquid, which is also shared by liquid $^4$He. The terms ``maxon'' and ``roton'' serve as convenient labels for the first maximum followed by a deep minimum of a longitudinal dispersion curve. We have found that in this domain and in the strong coupling regime, there is a close affinity between the dispersion curves of the classical 2DDS liquid and the low-temperature bosonic dipole liquid. More precisely, in a recent work \cite{19} we have shown that the dispersion curve of the classical 2DDS liquid, as generated by our MD simulations, falls in the narrow band of dispersion curves for the 2D bosonic dipole system: the band is bounded from above by the Feynman Ansatz (Eq.~\eqref{eq:1}) and from below by the CBF calculated dispersion. 

Thus, we believe that a detailed analysis of the collective mode behavior of the classical 2DDS liquid will reflect -- except for the damping -- the collective mode behavior of its quantum counterpart.  As to the damping, the high-$ka$ modes, in general, are strongly damped in a classical liquid to the extent that they become unobservable.  This seems to be the case in the 2D OCP and Yukawa plasmas. Here, we have found that the expected strong damping prevails in the 2DDS liquid as well. By contrast, the collective modes in the 2D quantum system in its superfluid phase are virtually undamped. Thus, we expect that the collective mode features that we analyze in the classical domain will correspond to the observable features in the collective modes of the 2D bosonic dipole system in its superfluid phase. We have considered three wave number domains for the analysis of the collective modes in the liquid phase: (i) the long-wavelength ($ka \to 0$) domain, where both longitudinal and transverse acoustic modes develop, (ii) the finite wave number domain spanning the maxon-roton portion of the longitudinal dispersion curve, and (iii) the high-$k$ domain where the dispersion is dominated by single-particle excitations. Our findings from the MD simulations are as follows: For wavenumbers $0\le ka\le 2.5$ extending somewhat beyond the maxon portion of the dispersion curve, the longitudinal collective excitations are robust over a wide range of coupling strengths $7\le \Gamma_D \le 60$. During the progressive deepening of the roton minimum with increasing coupling, its $ka \sim 3.7$ position remains more or less the same. In keeping what we have already stated, this wave number value in the classical 2DDS liquid is quite close to the location of the roton minimum in the quantum 2DDS at zero temperature \cite{8}. For  wave numbers well above the roton minimum, the dispersion assumes the character of single particle excitations.  This feature is also qualitatively the same -- albeit with different $k$-dependence --, as the one predicted and observed in the high-$k$ domain for the 2D cbosonic dipole system.

Additionally, we observe the existence of a faint maxon-maxon (M+M) harmonic frequency that persists over the broad range of wave numbers $0.05\le ka\le 5$. This feature can be related to the observation of roton-roton, roton-maxon, and maxon-maxon combination frequencies in the superfluid phase of $^4$He. Whether the absence of the first two combination frequencies is a feature that distinguishes the 2DDS from liquid $^4$He, due to the difference in interaction potentials, or these two combination frequencies are simply masked by classical noise is not clear.  More combination frequencies have been observed in the crystalline solid phase.  However, due to the anisotropy of the medium, the formation of combination frequencies in a lattice is a much more intricate matter than in an isotropic liquid. The definite identification of the frequencies with the maxon-maxon (M+M), maxon-roton (M+R and perhaps R-M), roton-roton (R+R) combinations would still require more study and further understanding of this process. 

We have compared the MD data with the theoretical analysis based on the QLCA approach. As expected, the range of $ka$ values marking agreement between QLCA theory and MD data increases with increasing coupling. At $0.05\le ka\le 2.5$ there is near-perfect agreement between theory and simulation. While, however, the QLCA dispersion closely emulates the qualitative features of the dispersion even at higher $ka$ values, in the neighborhood of the roton minimum it underestimates the roton frequency and overestimates its wave number. An improvement of the QLCA, with the capability to account for the strong angular correlations in the strong coupling domain may be needed to deliver even better agreement with MD observations.

\begin{acknowledgments}
This work has been partially supported by NSF Grants PHY-0812956, PHY-0813153, PHY-0903808, and by OTKA-PD-75113, OTKA-K-77653, MTA-NSF/102, and the Janos Bolyai Research Scholarship of the Hungarian Academy of Sciences.
\end{acknowledgments}

\providecommand{\noopsort}[1]{}\providecommand{\singleletter}[1]{#1}%

\end{document}